\documentclass[manuscript]{aastex}

\newcommand{\noprint}[1]{}

\begin{document}

\shorttitle{Optical Spectra of Flat-Spectrum ICRF Radio Sources}
\shortauthors{Titov et al.}

\title{Optical Spectra of Candidate  International
Celestial Reference Frame (ICRF) Flat-Spectrum Radio Sources}
\author{O.\ Titov}
\affil{Geoscience Australia, PO Box 378, Canberra, ACT 2601, Australia}
\email{oleg.titov@ga.gov.au}

\author{Laura M.\ Stanford}
\affil{Geoscience Australia, PO Box 378, Canberra, ACT 2601, Australia}

\author{Helen M.\ Johnston}
\affil{Sydney Institute for Astronomy, School of Physics, University of Sydney, NSW 2006, Australia}

\author{T.\ Pursimo}
\affil{Nordic Optical Telescope,
Nordic Optical Telescope Apartado 474E-38700 Santa Cruz de La Palma, Santa 
Cruz de Tenerife, Spain}

\author{Richard W.\ Hunstead}
\affil{Sydney Institute for Astronomy, School of Physics, University of Sydney, NSW 2006, Australia}

\author{David L.\ Jauncey}
\affil{CSIRO Astronomy and Space Science, ATNF \& Research School of Astronomy \& Astrophysics, Australian National University, Canberra, ACT 2611, Australia}

\author{K.\ Maslennikov}
\affil{The Central Astronomical Observatory at Pulkovo, Pulkovskoye Shosse, 65/1, 196140, St.Petersburg, Russia}

\and

\author{A.\ Boldycheva}
\affil{Ioffe Physical Technical Institute, 26 Polytekhnicheskaya, St. Petersburg, 194021, Russia}

\begin{abstract}

 Continuing our program of spectroscopic observations of ICRF sources, we
  present redshifts for 120 quasars and radio galaxies. Data were 
  obtained with
  five telescopes: the 3.58\,m ESO New Technology Telescope (NTT), 
  the two 8.2\,m Gemini telescopes,
  the 2.5\,m Nordic Optical Telescope (NOT), and the 6.0\,m Big Azimuthal
  Telescope (BTA) of the Special Astrophysical Observatory in Russia.

  The targets were selected from the International VLBI Service for
  Geodesy \& Astrometry (IVS) candidate International Celestial
  Reference Catalogue which forms part of an observational VLBI
  program to strengthen the celestial reference frame. We obtained
  spectra of the potential optical counterparts of more than 150
  compact flat-spectrum radio sources, and measured redshifts of 120
  emission-line objects, together with 19 BL Lac objects. These
  identifications add significantly to the precise radio--optical
  frame tie to be undertaken by {\it Gaia}, due to be launched in
  2013, and to the existing data available for analysing source proper
  motions over the celestial sphere.  We show that the distribution of
  redshifts for ICRF sources is consistent with the much larger sample
  drawn from FIRST and SDSS, implying that the ultra-compact VLBI
  sources are not distinguished from the overall radio-loud quasar
  population.

 In addition, we obtained NOT spectra for  five radio sources from the 
FIRST and NVSS catalogs, selected on the basis of their red colors, 
which yielded three quasars with $z>4$.

\end{abstract}


\section{Introduction}

The coming astrometric space mission, {\it Gaia}
\citep{Perryman2001,Mignard2012}, to be launched in 2013 by the
European Space Agency, will measure high precision positions
($\sim26\mu$as for $V=15$ mag, $\sim300\mu$as for $V=20$ mag,
\citep{debruijne}) and proper motions of $\sim$500,000 quasars
brighter than $m_{v} = 20$. This new optical astrometric catalog will
be linked with the current radio astrometric catalog, ICRF2, the
second realization of the International Celestial Reference Frame
\citep{Fey2009}. Optical counterparts of the extragalactic sources are
being sought to confirm their identification as quasars.
  
This paper is the second in our series aimed at investigating the
optical spectra of radio sources in the International VLBI Service for
Geodesy \& Astrometry (IVS) Reference Catalog; see
\citet{Schluter2007} for a description of the IVS history. Astrometric
Very Long Baseline Interferometry (VLBI) measures the differences in
arrival times of radio waves from ultra-compact, flat-spectrum radio
sources at telescopes positioned large distances apart.  This
procedure determines the positions of such sources to milliarcsecond
precision.
  
The International Celestial Reference System (ICRS) \citep{arias1995}
was adopted by the International Astronomical Union (IAU) as a
reference system with its origin at the barycentre of the Solar system
\citep{MacCarthy2004}, and axes fixed by the positions of selected
extragalactic radio sources.  The first realization of the
International Celestial Reference Frame (ICRF1) was used to establish
the orientation of the ICRS axes \citep{Ma1998}. The current
International Celestial Reference Frame, known as ICRF2, is based on a
catalog of 295 `defining' soures.  The formal weighted errors in the
radio positions are reported by \citet{Fey2009} to have an upper limit
to the noise floor of $41~\mu$as.

The IVS astrometric program has a total catalog of $>6000$ radio sources,
where $\sim 1200$ are observed on a regular basis.  In the Southern
Hemisphere, there is a significant deficit in candidate sources as well
as a lack of optical identifications.  By 2012 February, of the 3257  
objects with measured redshifts only 1213 are in the southern
hemisphere and only 287 have declinations south of $-40\degr$.  This   
paucity of redshifts in the south leads to problems in the analysis of   
apparent proper motions of the reference radio sources
\citep{Titov2009}.  To address this issue, an extensive program was
started in 2010 to find optical counterparts and determine redshifts
for southern IVS sources \citep{Titov2011}.

The quasars at high redshift ($z\geq2$) will be used for more intensive
observations at the VLBI facilities in the southern hemisphere.  VLBI
observations of weak sources will be undertaken with the 64-meter
telescope in Parkes, Australia, and two 26-meter telescopes in Hobart   
(Australia) and HartRAO (South Africa).  Stronger radio sources (flux
density $\geq$400 mJy) will be monitored with four 12-meter telescopes
recently installed in Australia: the AuScope network comprising Hobart,
Yarragadee and Katherine in Australia \citep{Titov2013} and Warkworth in
New Zealand. Several quasars found during the first observing run with the
NTT in 2010 August \citep{Titov2011} have now been tracked with the
AuScope radio telescopes in 2011--2012.

In this paper we continue our spectroscopic observations of the optical
counterparts of southern IVS sources, in particular those with a long VLBI
observational history. Some strong radio sources from the northern  
hemisphere have also been observed.  Optical identifications were sought
initially from the image and catalog data from the SuperCOSMOS Sky Surveys
on the grounds of their small digitization pixel scale and excellent  
astrometric accuracy \citep{Hambly2001}.  We also took advantage of the
Sloan Digital Sky Survey (SDSS; \citealt{York2000}) DR8 data release in
the regions where it was available.  This improved the identification
process, especially in regions of high stellar density.

In addition, we observed five weaker radio sources from the NVSS (NRAO VLA
Sky Survey; \citealt{Condon1998})  and FIRST catalogs (Faint Images of the
Radio Sky at Twenty cm; \citealt{Becker1995}) that we identified with
objects from the SDSS whose colors were typical of high redshift quasars.
Such high redshift quasars provide unique information about the early
stages of the Universe. The number of known radio sources with $z\geq4$ is
small, and we are exploiting this technique in an attempt to increase  
their number.

The observations and data reduction procedures are described in Section 2
and we report our results, along with detailed comments on individual
objects, in Section 3.

\section{Observations}

Spectroscopic observations were carried out at five optical facilities.

\noindent{\bf ESO NTT}: We had a 5-night observing run in Visitor Mode at
the European Southern Observatory (ESO) 3.58-meter New Technology
Telescope at La Silla in 2011 December (088.A-0021 (A)) using the ESO
Faint Object Spectrograph and Camera (EFOSC) system with grism \#13
covering the wavelength range 3685--9315~\AA. The seeing during
observations was typically $0\farcs{5} - 1\farcs{5}$, with a wavelength
resolution 21.2~\AA. Exposure times varied from 10 minutes to 1 hour
depending on the magnitude of each target and current sky conditions.  
Wavelength calibration was performed using the spectra of a HeNeAr
comparison lamp, resulting in an rms accuracy of 0.5~\AA.

\noindent{\bf Gemini}:  A large number of targets were observed in Service
Mode at the Gemini North and Gemini South 8.2-meter telescopes through the
Poor Weather Program (GS-2011A-Q-89, GN-2011B-Q-109, GS-2011A-Q-94) using
the Gemini Multi-Object Spectrograph (GMOS) system with grating R400. This
grating covers 4500~\AA\, centered either at 5200~\AA\ or 6500~\AA.  
As expected, the seeing and weather conditions were  variable but the 
program overall was very successful. The wavelength
resolution was $\sim$15~\AA, and an exposure time of 20 minutes was used
for all targets.  Wavelength calibration was performed using the spectra
of a CuAr lamp, resulting in an rms accuracy of $\sim$0.3~\AA.

\noindent{\bf NOT}:  Observations with the 2.5-meter Nordic Optical
Telescope on La Palma were carried out using the Andalucia Faint 
Object Spectrograph and Camera (ALFOSC) spectrograph,
either with grism \#5, or grism \#4 with the WG345 blocking filter. The
nominal wavelength range for grism \#4 is 3200--9100~\AA, with the
second-order blocking filter cutting below 3560~\AA. The red end of the
detector suffers fringing, so the effective long-wavelength limit is about
8000~\AA. For grism \#5 the nominal range is 5000--10250~\AA. The slit
width was $1\farcs{0}$\ or $1\farcs{3}$\, depending on the seeing. The
typical integration time was between 20 and 40 minutes. The longer
integrations were divided into two and the target was offset along the
slit in order to improve the fringe correction. For the single
integrations, internal halogen lamp images were taken before and after the
science frame.  Wavelength calibration was based on a HeNe lamp exposure
taken before the science frame(s), resulting in an rms accuracy of
$\sim$0.5~\AA.

\noindent{\bf BTA}:  Two objects were observed in Visitor Mode at the
6-meter Big Azimuthal Telescope (BTA) telescope, of the Special
Astrophysical Observatory in Russia in 2011 August, using the SCORPIO
multi-mode focal reducer with GR300 grism covering the wavelength range
3500--9500~\AA. The seeing during observations was about $2\arcsec$.  
Spectral resolution was typically 20~\AA.\\

Data reduction was performed with the {\sc iraf} software
suite\footnote{IRAF is distributed by the National Optical
  Astronomical Observatories, which are operated by the Association of
  Universities for Research in Astronomy, Inc., under contract to the
  National Science Foundation.} using standard procedures for spectral
analysis. We removed the bias and pixel-to-pixel gain variations from each
frame and then removed cosmic rays using the IRAF task {\sc szap}. Where
more than one exposure was obtained, the separate exposures were combined.  
Spectrum extraction, sky subtraction and wavelength calibration were then
carried out and the final one-dimensional spectra were flux-calibrated
with a spectrophotometric standard observed with the same instrumental
setup. Because the conditions were often non-photometric, especially for
observations made through the Gemini Poor Weather Program, the flux
calibration should be taken as approximate.

\section{Results}

Spectra of 120 IVS objects are shown in Fig.\ \ref{spectra}, along
with the line identifications.  A blue, dashed line indicates lines
that were used for redshift calculation, while a red, dot-dashed line
indicates lines that were detected, generally at a low signal-to-noise
ratio (S/N), but not used in determining the mean redshift.

Table\ \ref{emlines} lists the IVS sources with their ICRF2
coordinates (which refer to the epoch J2000.0), the telescope used for
each spectrum, the identified emission lines with their rest and
observed wavelengths, the mean redshift and error, and brief notes on
individual sources.  More detailed notes on individual sources
(indicated by an asterisk in the final column) are given in section
3.2.

The quoted errors $\Delta z$ in the mean redshift $\overline{z}$ are 
given by

\( \Delta z = \{[(\sigma_z)^2 + 
(\Delta\lambda/\overline{\lambda_0})^2]/N\}^{1/2},  \)

\noindent where $\sigma_z$ is the measured standard deviation among
the independent estimates of $z$, $\Delta\lambda$ is the rms error in
the wavelength calibration (typically 0.5\ \AA), and
$\overline{\lambda_0}$ is the mean rest wavelength of the $N$ lines
used to measure $\overline{z}$.  Single-line redshifts (mostly
\ion{Mg}{2}) are assigned a conservative error of 0.001 if the
signal-to-noise ratio is high and the line is symmetric.  If the
signal-to-noise ratio is low or the line is broad or asymmetric an
(arbitrary) error of 0.002 has been assigned; in two extreme cases
where the signal-to-noise ratio is low {\it and\/} the line is broad
or asymmetric (IVS B0633$-$26B and B1129$-$161) the redshift is given
with a colon(:) appended and no error.

Nineteen objects, listed in Table\ \ref{BLL} with their ICRF2
positions \citep{Fey2009}, were found to have a good signal-to-noise
ratio (typically S/N$\sim60-110$) but featureless spectra and hence
are identified as probable BL Lac objects.  Their spectra are shown in
Fig.\ \ref{BLLspectra}.


A further 18 targets had spectra with a signal-to-noise ratio that was too 
low for confident spectral classification; these are listed in Table\ 
\ref{fa} with their ICRF2 positions.

There were five IVS targets that returned stellar spectra.  This was
assumed to be the result of foreground obscuration and, in most cases,
small but significant offsets between SuperCOSMOS optical and ICRF2 radio
positions.  In two cases the correct identification was found when the
fields were reobserved in excellent seeing.  Further discussion of these
five objects is given in Section\ \ref{close_objects}.

\setcounter{figure}{2}

\subsection{Separation of close objects\label{close_objects}}

Occasionally, Galactic stars are found close on the sky to the
radio position, leading to possible misidentification. Here we
note several cases that were encountered during the observing runs.

\begin{figure}[ht]
\centering
\includegraphics[height=3.5cm]{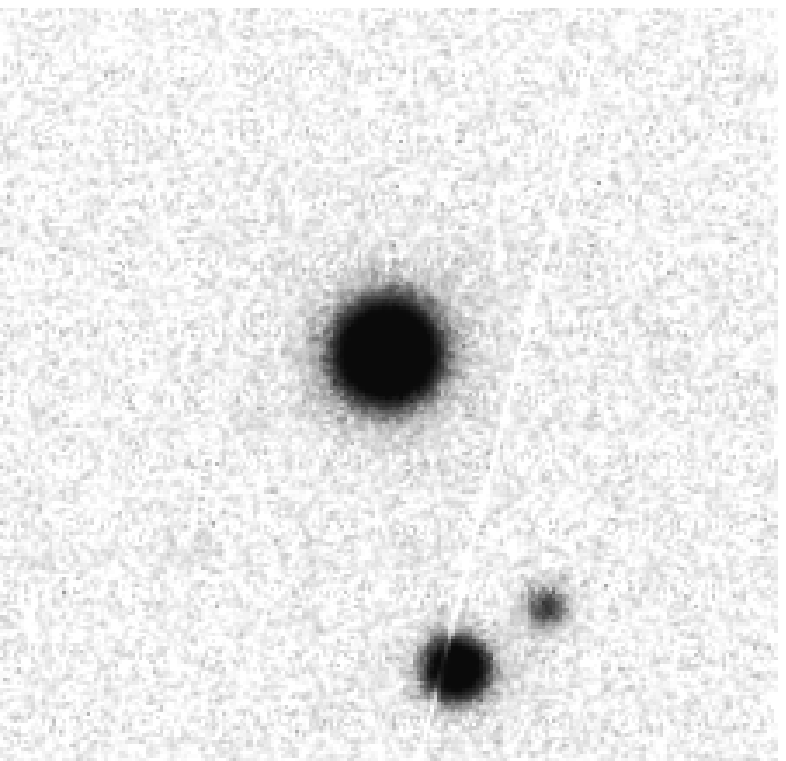}%
\hspace{3mm}
\includegraphics[height=3.5cm]{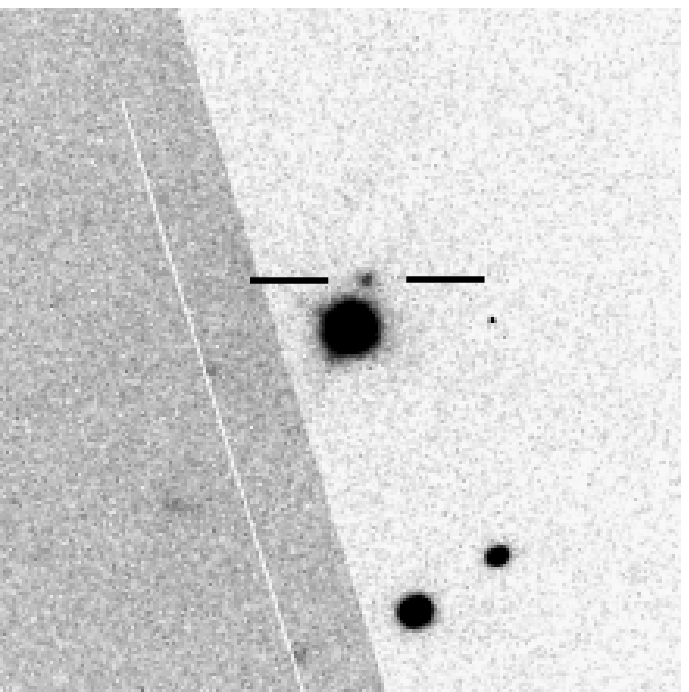}
 \caption{Two acquisition images of the sky field around the quasar IVS
B2300$-$307 made at the NTT in 2010 August (left: seeing $2''$) and 2011
December (right: seeing $0\farcs{6}$). Each image is 50$''$ on a side;  
north is up and east to the left.  The foreground star in the centre of
the left image completely obscures the quasar due to its large seeing
disk. In the right image the much better seeing reveals the faint quasar
($R\sim$22, marked between the bars), separated by $3\farcs{7}$ from the
star.  (The diagonal step in contrast in the right image is an artefact of
fast readout using two amplifiers for the acquisition image.)}
 \label{2300images}
 \end{figure}

\begin{itemize}

\item IVS B0900$-$664---the spectrum obtained was that of a red M star,
which is offset by $0\farcs{52}$ from the ICRF position.  A faint 
counterpart was seen in the NTT B-band acquisition image, but a much 
longer integration in good seeing will be needed to secure a redshift.


\item IVS B0905$-$202---the nearest object, as seen by SuperCOSMOS, was a
$R=14.6^{\rm m}$ stellar object located $1\farcs{55}$ from the radio
position. Our NTT acquisition image, taken in $0\farcs{6}$ seeing, showed
a faint object at the radio position, but on the limb of the stellar disk.  
Spectroscopy of this faint $R\sim 22$ object did not reveal any clear
emission lines.

\item IVS B1657$-$261---located in a very crowded star field in the
Galactic bulge at a latitude of $b=9\fdg{7}$. Identification of the
optical counterpart may not be feasible.

\item IVS B1946$-$582---based on the SuperCOSMOS optical position,
 the radio minus optical position difference is only $0\farcs{06}$ in
right ascension and $-0\farcs{04}$ in declination, but the Gemini South
observation of this $R\sim 19$ object showed a typical stellar spectrum.
The radio source has a flat spectrum with flux density 200--300 mJy at cm
wavelengths.

\item IVS B2300$-$307---the obscuration of the optical field of this
source by a foreground star was discussed previously \citep{Titov2011}.
However, a later 80\,s image in $0\farcs{6}$ seeing (Fig.\
\ref{2300images})  revealed a faint $R\sim 22$ object coincident with the
radio position. Our spectroscopy yielded a redshift of $1.039\pm 0.002$
based on weak \ion{C}{3}] and \ion{Mg}{2} emission.

\end{itemize}


\subsection{Notes on individual targets}

\begin{itemize}



\item IVS B0233$-$478---the broad feature attributed to \ion{Al}{3} and
\ion{C}{3}] may also include \ion{Si}{3}~$\lambda$1890.

\item IVS B0417$-$302---prominent \ion{Fe}{2} multiplet emission; 
intervening \ion{Mg}{2} doublet absorption at $z_{\rm abs}=0.8395 \pm 
0.0005$.

\item IVS B0447$-$507---possible double-peaked \ion{C}{3}] line and
prominent \ion{Fe}{2} multiplet bands.

\item IVS B0448$-$482---poor redshift match to \ion{C}{3}] may indicate
the presence of \ion{Si}{3}~$\lambda$1890; prominent \ion{Fe}{2} multiplet
bands.
  

\item IVS B0521$-$262---strong associated absorption in the Ly\,$\alpha$
emission line and a clear detection of the Lyman limit at the emission
redshift.

%



\item IVS B0529$-$031---faint optical counterpart with a very red
spectrum, showing a single broad emission line identified as \ion{Mg}{2}
with associated \ion{Mg}{2} absorption at $z_{\rm abs}=1.8744 \pm 0.0004$,
corresponding to a relative blue-shift of $\sim$1500\,km\,s$^{-1}$; a
further absorption line, at 8182.1\,\AA\ remains unidentified.  Given the
low signal blueward of 6000\,\AA, together with the faint SDSS blue
magnitudes ($u= 24.41,\ g=21.93$), it is not surprising that \ion{C}{3}]
and \ion{C}{4} are not detected.

\item IVS B0548$-$527---redshift in Table\ \ref{emlines} is for the
forbidden lines;  the permitted lines of \ion{Mg}{2} and H\,$\beta$ show a
redshift systematically higher by $830\pm 80$\,km\,s$^{-1}$.

\item IVS B0554+242---narrow self-absorption in Ly\,$\alpha$ and
\ion{C}{4} at $z_{\rm abs}=3.2319 \pm 0.0010$.



\item IVS B0608$-$230---wavelengths for \ion{C}{3}] and \ion{C}{4} are not
consistent, suggesting the presence of additional lines or asymmetric
structure; in addition, the Ly\,$\alpha$ wavelength is likely to be
affected by absorption in the blue wing.



\item IVS B0633$-$26B---a very faint galaxy; a low signal-to-noise ratio
spectrum, with a single broad emission line, assumed to be \ion{Mg}{2}.


\item IVS B0810$-$180---very broad line wings in \ion{C}{4} and
Ly\,$\alpha$, extending $\sim$30,000\,km\,s$^{-1}$ redward of the line
peaks.

\item IVS B0828$-$064---poor consistency in redshift between
  \ion{C}{4} and \ion{C}{3}], possibly due to associated absorption in
  the blue wing of \ion{C}{4}.  \ion{Mg}{2} is present with a low
  signal-to-noise ratio just redward of the atmospheric A-band.



\item IVS B0844$-$557---redshift in Table\ \ref{emlines} is for the
forbidden lines; permitted lines \ion{Mg}{2} and H\,$\beta$ are displaced
$\sim$1000\,km\,s$^{-1}$ to higher redshift.



\item IVS B0948$-$860---single emission line, assumed to be Ly\,$\alpha$;
the adopted redshift, $z=3.696$, is consistent with the 40\% continuum
depression blueward of the emission line due to the Ly\,$\alpha$ forest,
and the possible detection of \ion{C}{4} at the red edge of the 
spectrum.

\item IVS B0952$-$185---associated absorption at the emission redshift in 
both Ly\,$\alpha$ and \ion{C}{4}.

\item IVS B0956$-$409---Ly\,$\alpha$ is strongly self-absorbed.

\item IVS B1004$-$125---reported as having $z=0.24$ in Simbad (no
reference given), clearly inconsistent with our Gemini spectrum; strong
intervening heavy-element system at $z_{\rm abs}=1.5786 \pm 0.0002$ based 
on
\ion{Fe}{2}\,$\lambda\lambda$2344, 2382, 2586, 2600 and
\ion{Mg}{2}\,$\lambda\lambda$2796, 2803 absorption.



\item IVS B1020+270---extended red wing in \ion{C}{4}; \ion{C}{3}] is not
detected due to strong CCD fringing.

\item IVS B1039$-$474---\ion{Si}{4}, \ion{C}{4} and \ion{C}{3}] emission
lines all show strongly extended blue wings; redshift is based on the peak
positions of the lines.



\item IVS B1127$-$443---single-line redshift is supported by stellar 
absorption features (G-band, \ion{Mg}{1}b) noted in the plotted 
spectrum.

\item IVS B1143$-$696---spectrum is very similar to that of 3C\,273, with 
strong \ion{Fe}{2} multiplet emission in the region around H\,$\beta$ 
\citep{wampler1967}. 



\item IVS B1722+562---strong associated absorption system at $z_{\rm
abs}=2.2463 \pm 0.0002$ seen in \ion{C}{4}, \ion{Si}{4}, \ion{N}{5} and
Ly\,$\alpha$.  This has led to a relatively large uncertainty in emission 
redshift.




\item IVS B2235$-$556---strong self-absorption in the red wing of 
\ion{C}{4}.



\item IVS B2334$-$525---broad \ion{Mg}{2}, possibly double-peaked with an
extended red wing.


\item IVS B2341+295---very blue object based on POSS\,II sky survey
images; rise in the spectrum redward of 7000\,\AA\ may be due to the
underlying galaxy or an unrelated object on the spectrograph slit.  
Intervening \ion{Mg}{2} absorber at $z=0.8644 \pm 0.0004$.


\end{itemize}

\subsection{Spectra of  color-selected quasars}

\begin{figure}[ht]
\epsscale{0.8}
\plotone{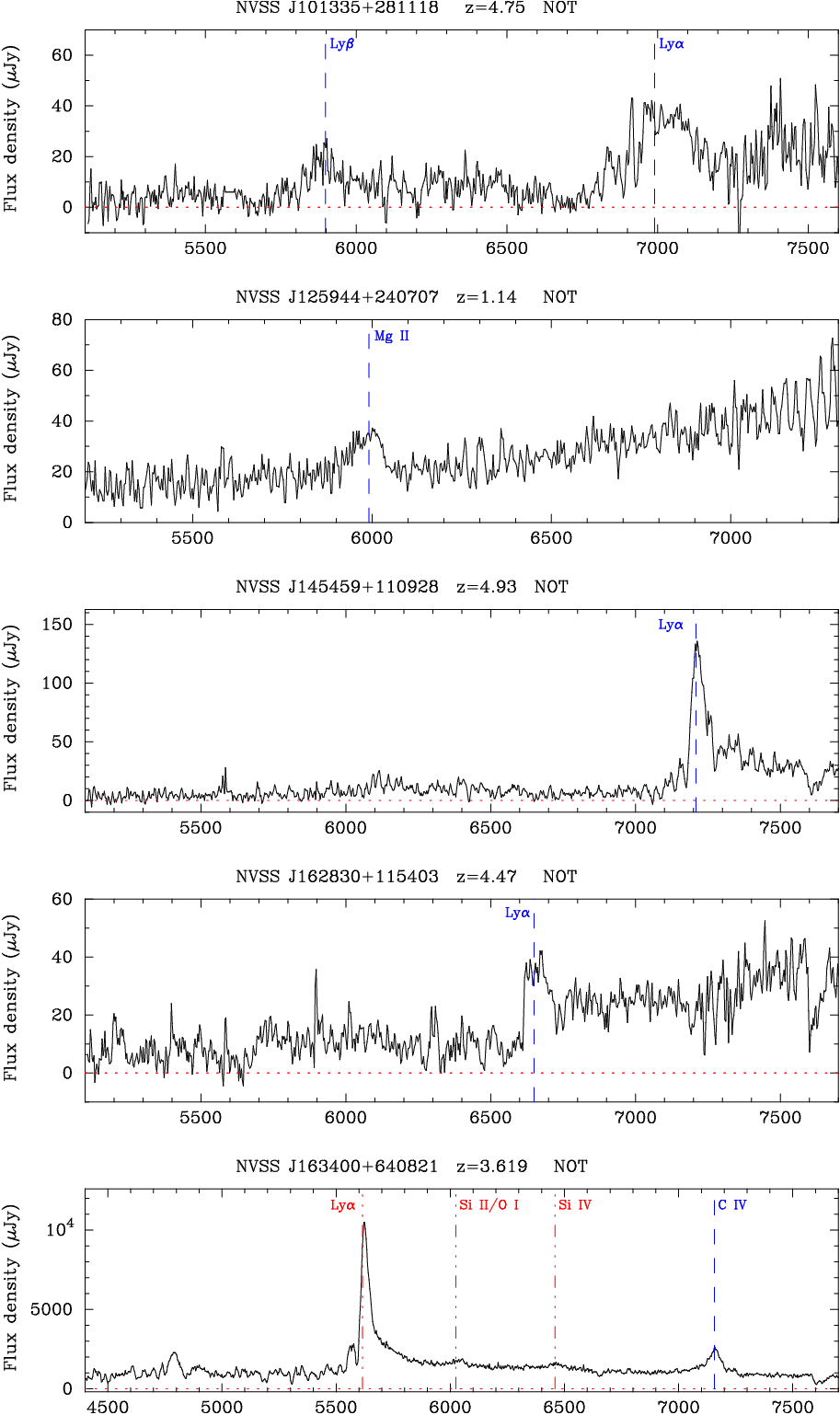}
\caption{Spectra of the five color-selected quasars from NVSS/FIRST
  and SDSS.  Dashed lines (blue) indicate emission lines used for
  redshift determination; dot-dashed lines (red) indicate lines
  detected at a lower signal-to-noise ratio or blended.  Further details are
  given in Table\ \ref{nvss_red}.\label{nvss}}
\end{figure}

In an attempt to find more high-redshift radio quasars for our VLBI proper
motion studies as a function of redshift, we selected five weak radio
sources from the FIRST and NVSS catalogs for which the SDSS colors
suggested the likelihood of $u$- or $g$-band dropouts and the possibility
of high redshifts.  Spectra were obtained at the NOT on the nights of 2012
May 24--26.  The SDSS colors, wavelengths and redshifts for the five
color-selected radio quasars are given in Table\ \ref{nvss_red}. The
`bluest' of the five, NVSS J125944+240707, was a $z = 1.139$ radio galaxy,
but the other four proved to be broad emission-line quasars with redshifts
in excess of 3.5, including three with $z>4$. This is an excellent return
for our search and a striking result from the NOT, the smallest of the
telescopes used in this program.  The five spectra are shown in Fig.\
\ref{nvss} and further information is given in Table\ \ref{nvss_red}.  
Redshifts for the  $z>4$ quasars are based  on estimated 
wavelengths for Ly$\alpha$ (and, in one case, Ly$\beta$) and 
are very uncertain because of strong absorption in the 
blue wing of the line.

A comparison between the FIRST and NVSS flux densities at 1.4\,GHz for the
five sources showed that one, NVSS J145459+110928, appeared to show
evidence of variability, with $S_{\rm NVSS}=9.8 \pm 0.5$ mJy and $S_{\rm
FIRST}= 15.07\pm 0.14$ mJy. This is also the only source that is 
unresolved in FIRST.  The other four sources all show minor extension at 
the $1\arcsec$ level.

\section{Redshift distributions}

Together with the 31 redshifts reported in \cite{Titov2011}, we have
now accumulated over 150 redshifts for IVS sources, mostly in the
south.  Their optical counterparts are systematically fainter than
those in the \cite{Titov2009} compilation.  Since the IVS selection
process is directed at compact, milliarcsecond radio quasars, it is
important to test whether their redshift distribution differs from
that of the FIRST-SDSS quasar sample \citep{kimball2011}, which is
selected without regard to morphology, spectral index or angular size
and extends to much lower flux densities.

\begin{figure*}[htb]
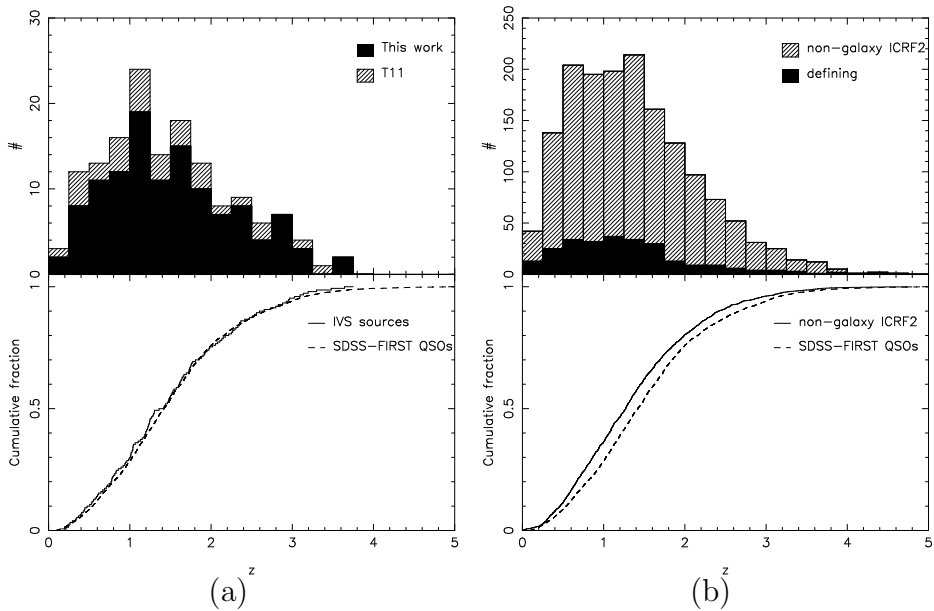

\centering
\includegraphics[angle=-90,width=6cm,clip]{Fig5A.ps}%
\hspace*{3mm}%
\includegraphics[angle=-90,width=6cm,clip]{Fig5B.ps}\\
 (a) \hspace*{56mm} (b)
 \caption{(a) Distribution of redshifts among the $\sim$150 IVS sources
from this paper and the previous paper \citep[abbreviated above as
T11]{Titov2011}.  The lower panel shows the cumulative redshift
distribution (K-S test) compared with the FIRST-SDSS sample
\citep{kimball2011}.  (b) Corresponding redshift distribution and K-S test
for 1594 ICRF2 quasars from
 \cite{Titov2009}; see text.
\label{zdist}}
 \end{figure*}

The redshift distribution of the sources from this paper and
\cite{Titov2011} is shown in Figure~\ref{zdist}(a).  For comparison,
Figure~\ref{zdist}(b) shows the distribution of 1594 point sources
with known redshifts from ICRF2, using redshift data from
\cite{Titov2009}.  The SuperCOSMOS \cite{Hambly2001} morphological
classification was used to remove galaxies from the \cite{Titov2009}
list, and the sample was further restricted to those with
radio-optical offsets $<1''$ and Galactic latitudes $>10^{\circ}$
(Schaefer et al.\ 2013, in prep.).


The redshift distribution for the sources from this paper and the earlier
paper \citep{Titov2011} is completely consistent with having been drawn
from the same redshift distribution as FIRST-SDSS quasars
\citep{kimball2011}, with a Kolmogorov-Smirnov test giving a probability
$p=0.84$ of the two samples being drawn from the same distribution.  Thus
the sharp drop-off that we see in the number of quasars at $z \gtrsim 2$\
is a reflection of the underlying quasar distribution, and not a selection
effect of the IVS candidate sources.  While the K-S plot for the ICRF2
sample shows an apparent excess of low-redshift sources compared with
FIRST-SDSS, the difference is not significant, with a K-S probability of
$p=0.24$.  We attribute this excess to selection effects arising from the
redshifts in \citep{Titov2009} being drawn from the literature, and
therefore likely to be biased towards brighter, lower-redshift quasars.

\section{Summary and conclusion}

We present redshifts and spectra for 120 emission-line objects
identified with radio sources from the candidate International
Celestial Reference Catalog.  Most of the target objects are in the
south and many had not previously been optically identified.  While
redshifts were usually based on two or more lines, those for 22
objects were based on a single emission line, in most cases assumed to
be \ion{Mg}{2}; many of these single-line redshifts were supported by
other spectral information and most are considered reliable.  In
addition, we classed 19 sources as probable BL Lac objects, based on
a high signal-to-noise ratio but featureless spectra.  A further 18
targets were considered to have a signal-to-noise ratio too low for
confident spectral classification.

The distribution of redshifts from this paper, together with those from
our earlier paper \citep{Titov2011}, is consistent with the much larger
sample drawn from FIRST and SDSS \citep{kimball2011}.  This implies that
the ultra-compact, flat-spectrum sources that make up the IVS Reference 
Catalog are not distinguished from the radio quasar population at large.  
On the other hand, the distribution of redshifts from the much larger 
sample drawn from ICRF2 \citep{Titov2009} has a small, but not significant 
excess of low-redshift quasars, almost certainly the result of 
observational selection.

\section{Acknowledgments}

This paper is based on observations collected at five  telescopes:

1. ESO New Technology Telescope, under the European
Organisation for Astronomical Research in the Southern Hemisphere,
Chile under program 088.A-0021(A).

2. Two Gemini Observatories, which are operated by the Association of
Universities for Research in Astronomy, Inc., under a cooperative
agreement with the National Science Foundation on behalf of the Gemini
partnership: the National Science Foundation (United States), the
Science and Technology Facilities Council (United Kingdom), the
National Research Council (Canada), CONICYT (Chile), the Australian
Research Council (Australia), Ministurio da Ciuncia, Tecnologia e
Inovacio (Brazil) and Ministerio de Ciencia, Tecnologia e Innovacion
Productiva (Argentina) under programs GS-2011A-Q-89 and GS-2011B-Q-94
(Gemini South), and GN-2011B-Q-109 (Gemini North).

3. Six-meter Big Azimuthal Telescope (BTA) operated by the
Special Astrophysical Observatory (Russia)

4. Nordic Optical Telescope, operated on the island of
La Palma jointly by Denmark, Finland, Iceland, Norway and Sweden,
in the Spanish Observatorio del Roque de los Muchachos of the
Instituto de Astrofisica de Canarias.


Two of us, Titov and Jauncey, were supported by a travel grant from
the Australian Nuclear Science Technology Organisation (ANSTO) in
their Access to Major Research Facilities Program (AMRFP) (reference
number AMRFP 10/11-O-31) to travel to the BTA telescope in Russia.

Funding for SDSS-III has been provided by the Alfred P. Sloan Foundation,
the Participating Institutions, the National Science Foundation and the
U.S. Department of Energy Office of Science. The SDSS-III web site is
http://www.sdss3.org/.

SDSS-III is managed by the Astrophysical Research Consortium for the
Participating Institutions of the SDSS-III Collaboration including
the University of Arizona, the Brazilian Participation Group,
Brookhaven National Laboratory, University of Cambridge, Carnegie
Mellon University, University of Florida, the French Participation
Group, the German Participation Group, Harvard University, the
Instituto de Astrofisica de Canarias, the Michigan State/Notre
Dame/JINA Participation Group, Johns Hopkins University, Lawrence
Berkeley National Laboratory, Max Planck Institute for Astrophysics,
Max Planck Institute for Extraterrestrial Physics, New Mexico State
University, New York University, Ohio State University, Pennsylvania
State University, University of Portsmouth, Princeton University,
the Spanish Participation Group, University of Tokyo, University of
Utah, Vanderbilt University, University of Virginia, University of
Washington, and Yale University.

AuScope is funded under the National Collaborative Research
Infrastructure Strategy (NCRIS), an Australian Commonwealth Government
Programme.

This paper is published with the permission of the CEO, Geoscience Australia.\\

\copyright\ Commonwealth of Australia (Geoscience Australia) 2013. This product is released under the Creative Commons Attribution 3.0 Australia Licence. http://creativecommons.org/licenses/by/3.0/au/deed.en


\clearpage

\begin{deluxetable}{lllclcclp{23mm}}
\tabletypesize{\scriptsize}
\renewcommand\arraystretch{0.8}
\setlength{\tabcolsep}{4pt}
\tablecolumns{5}
\tablewidth{0pc}
\tablecaption{Observed emission lines \label{emlines}}
\tablehead{
\colhead{Source}    &
\colhead{RA (J2000)\tablenotemark{a}} & \colhead{Dec (J2000)\tablenotemark{a}} &
\colhead{Telescope\tablenotemark{b}} & \colhead{Line} &
\colhead{Rest $\lambda$}   & \colhead{Obs $\lambda$\tablenotemark{c}} &
\colhead{$\overline{z}$} & 
\colhead{Note\tablenotemark{d}}
}
\startdata
IVS B0022$-$227 & 00 25 24.2474 & $ -$22 27 47.596 & GS & Mg {\sc ii}    &  2799.9  & 5136.3   &$0.834\pm0.001$  & S \\
& & & & & \\
IVS B0032$+$276 & 00 34 43.4861 & $ +$27 54 25.721 & BTA & Ly $\alpha$   &  1215.7  & 4820.4   &                 & \\
                &               &                  &     & N {\sc v}     &  1240.1  & [4921.7]  &                 & \\
                &               &                  &     & Si {\sc iv}   &  1398.3  & [5545.9]  &                 & \\
                &               &                  &     & C {\sc iv}    &  1549.5  & 6141.2   &$2.9642\pm0.0009$  & \\
& & & & & \\
IVS B0044$-$846 & 00 44 26.6883 & $ -$84 22 39.988 & GS & Mg {\sc ii}   & 
2799.9   & 5697.2   &$1.035\pm0.001$   & S $z=1.032$\ [S12] \\
\enddata
\tablenotetext{a}{ICRF2 radio position}
\tablenotetext{b}{Telescope abbreviations: BTA = Big Azimuthal
Telescope, GS = Gemini North, NOT = Nordic Optical Telescope, NTT =
ESO New Technology Telescope}
\tablenotetext{c}{Square brackets indicate the line was not used for redshift determination}
\tablenotetext{d}{``S'' denotes a single-line redshift; * indicates a
note in Section~3.2. Previously determined redshifts are from: [A09]
\citet{afanas2009}; [S12] \citet{Shaw2012}; [S] Simbad database; [6dF] 6dF survey, \citet{Jones2009}}
\end{deluxetable}

\begin{deluxetable}{llllc}
\tablecaption{Probable BL Lac objects\label{BLL}}
\tablewidth{0pt}
\tablehead{
  \colhead{Source} &
\colhead{RA (J2000)\tablenotemark{a}} & \colhead{Dec
(J2000)\tablenotemark{a}} & Telescope}
\startdata
IVS B0026$-$710 & 00 28 41.5626 &  $-$70 45 15.926  &GS      \\
IVS B0313$-$171 & 03 15 27.6783 &  $-$16 56 29.711  &NTT      \\
IVS B0608$-$187 & 06 10 17.8861 &  $-$18 47 40.090  &GS     \\
IVS B0633$-$263 & 06 35 20.9090 &  $-$26 20 39.866  &NTT     \\
IVS B0700$-$661 & 07 00 31.2531 &  $-$66 10 45.232  &NTT     \\
IVS B0729$-$826 & 07 23 39.4063 &  $-$82 47 41.089  &NTT     \\
IVS B0744$-$158 & 07 46 18.2360 &  $-$15 55 34.745  &NTT     \\
IVS B0752$-$116 & 07 54 26.4564 &  $-$11 47 16.948  &GS     \\
\enddata
\tablenotetext{a}{ICRF2 radio position}
\end{deluxetable}

\begin{deluxetable}{llllc}
\tablecaption{Objects with low signal-to-noise spectra\label{fa}}
\tablewidth{0pt}
\tablehead{
  \colhead{Source} &
\colhead{RA (J2000)\tablenotemark{a}} & \colhead{Dec
(J2000)\tablenotemark{a}} & Telescope }
\startdata
IVS B0205$-$242 & 02 07 33.3992 &  $-$24 02 02.210  &NTT     \\
IVS B0227$-$542 & 02 29 12.7856 &  $-$54 03 24.036  &NTT     \\
IVS B0244$-$050 & 02 46 49.1315 &  $-$04 51 10.933  &NTT     \\
IVS B0253$-$754 & 02 53 10.8816 &  $-$75 13 16.686  &GS     \\
IVS B0334$-$546 & 03 35 53.9248 &  $-$54 30 25.114  &NTT     \\
IVS B0458+138   & 05 01 45.2708 &  +13 56 07.220    &NTT   \\
IVS B0516$-$621 & 05 16 44.9261 &  $-$62 07 05.389  &NTT   \\
IVS B0600+219   & 06 03 51.5570 &  +21 59 37.698    &GN     \\
IVS B0628$-$627 & 06 28 57.4877 &  $-$62 48 44.744  &NTT     \\
IVS B0705$-$649 & 07 05 26.0526 &  $-$65 00 43.484  &NTT     \\
IVS B0718$-$523 & 07 19 14.2919 &  $-$52 28 32.506  &NTT     \\
IVS B0821$-$094 & 08 23 52.0233 &  $-$09 39 25.907  &NTT     \\
IVS B0905$-$202 & 09 07 54.0404 &  $-$20 26 49.475  &GS  \\
IVS B0939$-$077 & 09 42 21.4614 &  $-$07 59 53.204  &GS     \\
IVS B1118+073   & 11 20 38.4438 &  +07 04 47.174    &NTT   \\
IVS B1134$-$739 & 11 36 09.6597 &  $-$74 15 45.274  &NTT     \\
IVS B1821$-$525 & 18 25 13.8097 &  $-$52 30 58.281  &GS     \\
IVS B2020$-$015 & 20 23 32.8163 &  $-$01 23 42.153  &GS      \\

\enddata
\tablenotetext{a}{ICRF2 radio position}
\end{deluxetable}

\begin{deluxetable}{llllllllllc}
\tabletypesize{\scriptsize}
\renewcommand\arraystretch{0.8}
\setlength{\tabcolsep}{4pt}
\tablewidth{0pt}
 \tablecaption{SDSS magnitudes and redshifts for five 
color-selected NVSS radio sources \label{nvss_red}}
\tablehead{
\colhead{Source}    &
\colhead{RA (J2000)\tablenotemark{a}} & \colhead{Dec (J2000)\tablenotemark{a}} &
\colhead{$g$} & \colhead{$r$} & \colhead{Telescope\tablenotemark{b}} &
\colhead{Line} & \colhead{Rest $\lambda$}   & \colhead{Obs $\lambda$\tablenotemark{c}} &
\colhead{z\tablenotemark{d}}
}
\startdata
NVSS J101335+281118 &  10 13 35.49 & +28 11 18.3 & 25.76 & 22.03   & NOT & Ly\,$\alpha$  & 1215.7 & 6990  &   \\  
                    &              &             &      &        &     & Ly\,$\beta$     &1025.7  & 5897 &  4.75: \\
\\
NVSS J125944+240707 &  12 59 44.59 & +24 07 07.9 & 22.94  & 21.82    & NOT & Mg\,{\sc ii}   & 2799.9 & 5895.5 & $1.138 \pm 0.001$ \\
\\
NVSS J145459+110928 &  14 54 59.36 & +11 09 28.5 & 26.09 & 22.67   & NOT & Ly\,$\alpha$  & 1215.7 & 7208  &  4.93:    \\  
\\
NVSS J162830+115403 &  16 28 30.55 & +11 54 03.5 & 25.27 & 22.25   & NOT & Ly\,$\alpha$  & 1215.7 & 6650  &  4.47:    \\
\enddata 
\tablenotetext{a}{NVSS radio position}
\tablenotetext{b}{Nordic Optical Telescope}
\tablenotetext{c}{Square brackets indicate the line was not used for redshift determination}
\tablenotetext{d}{Colon indicates approximate redshift based on Ly\,$\alpha$ alone}
\tablenotetext{e}{Magnitudes are $u$ and $g$, respectively, for this object} 
\end{deluxetable}

\begin{figure*}
\centering
\figurenum{1}
\includegraphics[width=0.85\textwidth]{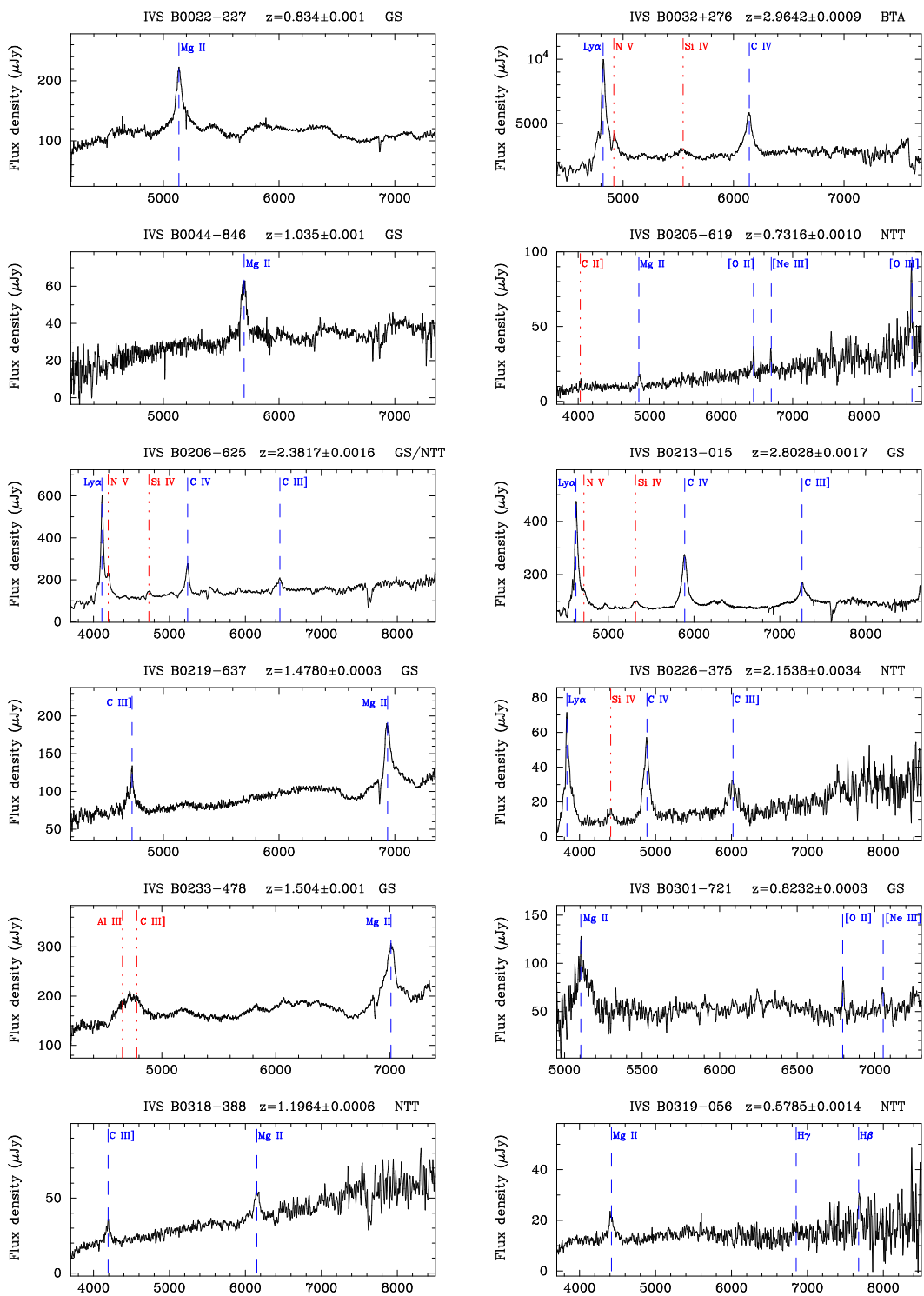}
 \caption{Optical spectra for 120 emission-line IVS targets. Dashed lines
(blue) indicate emission lines used for redshift determination; dot-dashed
lines (red) indicate lines detected at a lower
signal-to-noise ratio or blended.\label{spectra}}
 \end{figure*}

\begin{figure*}
\centering
\figurenum{1 (continued)}
\includegraphics[width=0.85\textwidth]{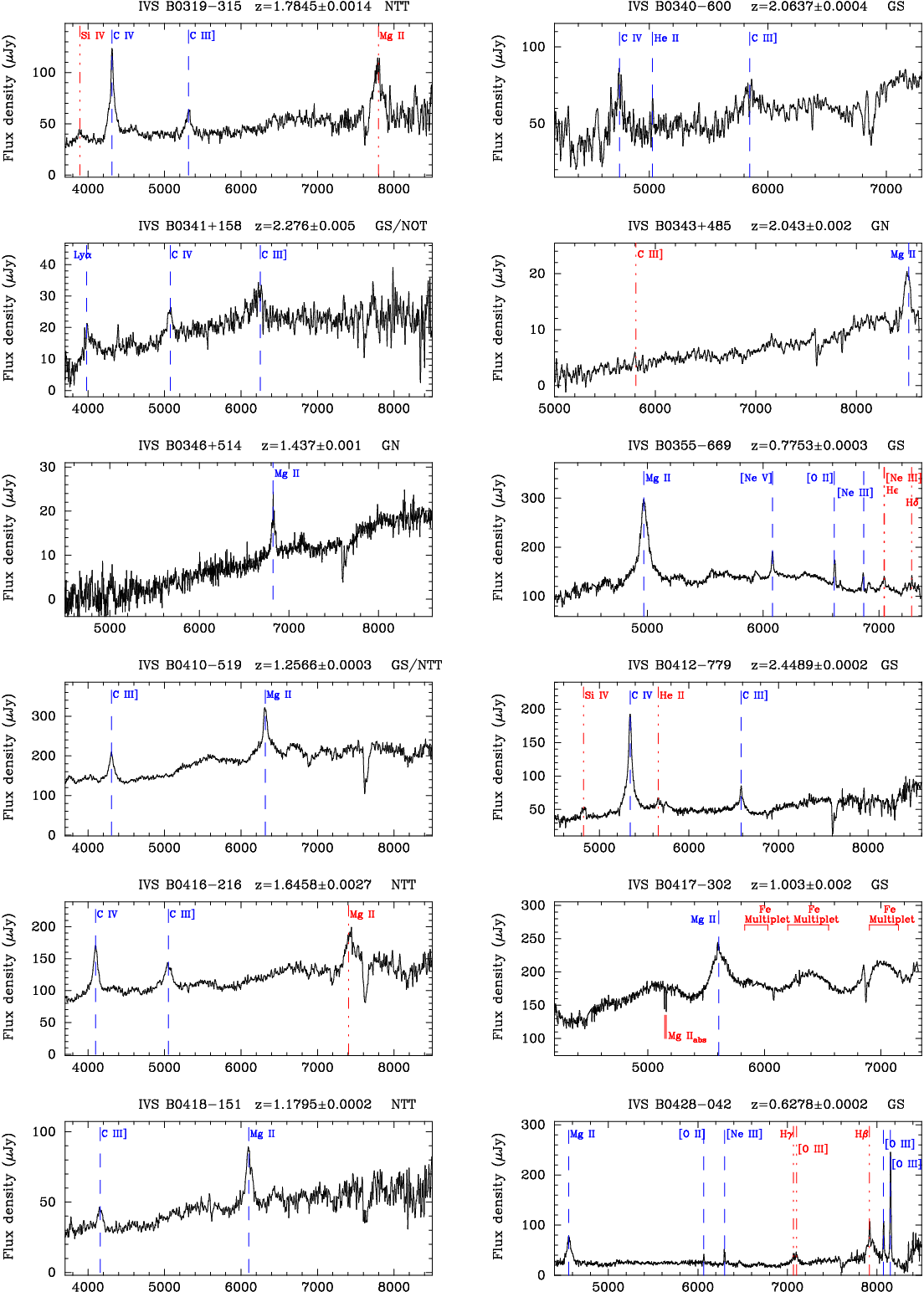}
\caption{}
\end{figure*}

\begin{figure*}
\centering
\figurenum{1 (continued)}
\includegraphics[width=0.85\textwidth]{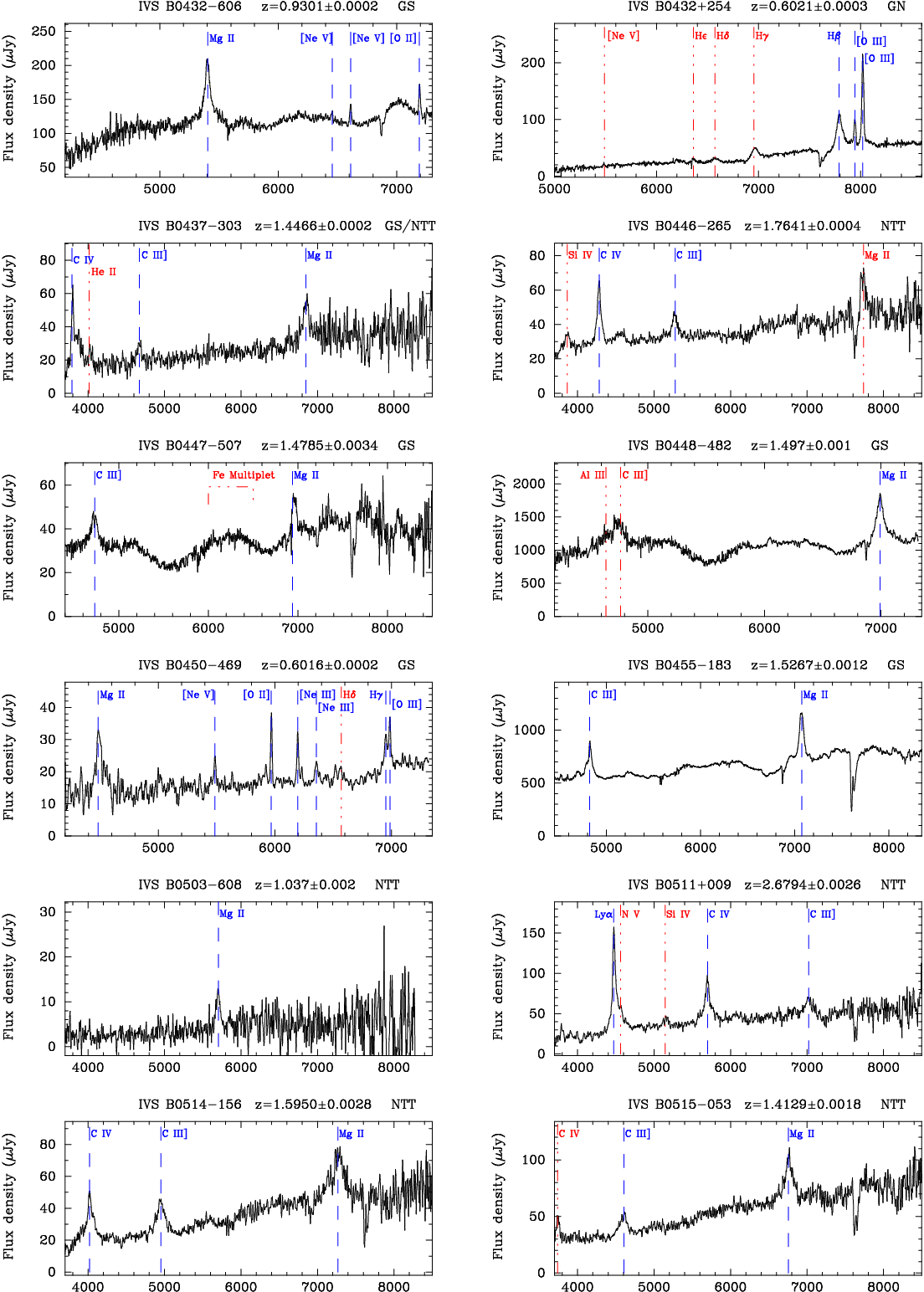}
\caption{}
\end{figure*}

\begin{figure*}
\centering
\figurenum{1 (continued)}
\includegraphics[width=0.85\textwidth]{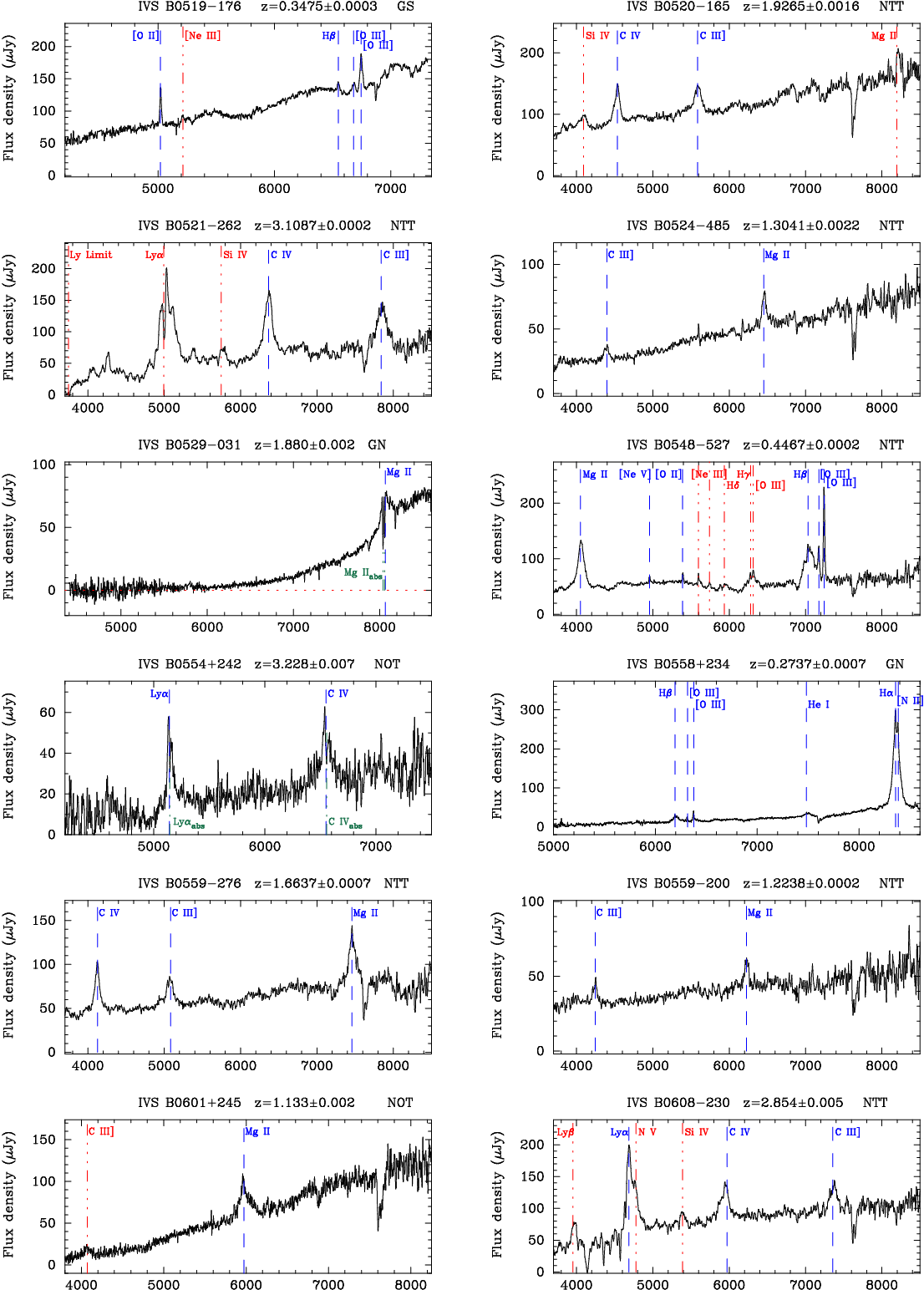}
\caption{}
\end{figure*}

\begin{figure*}
\centering
\figurenum{1 (continued)}
\includegraphics[width=0.85\textwidth]{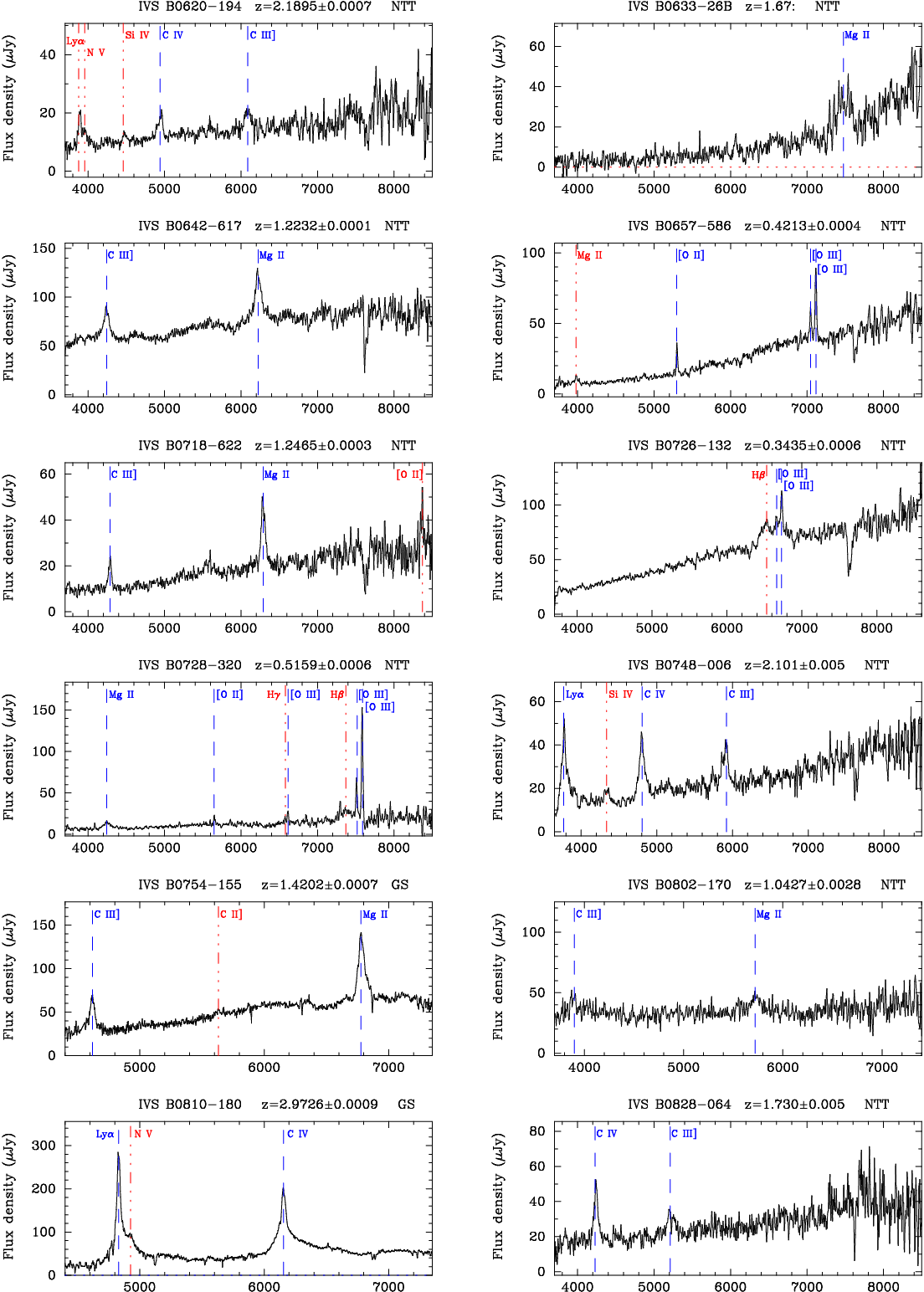}
\caption{}
\end{figure*}

\begin{figure*}
\centering
\figurenum{1 (continued)}
\includegraphics[width=0.85\textwidth]{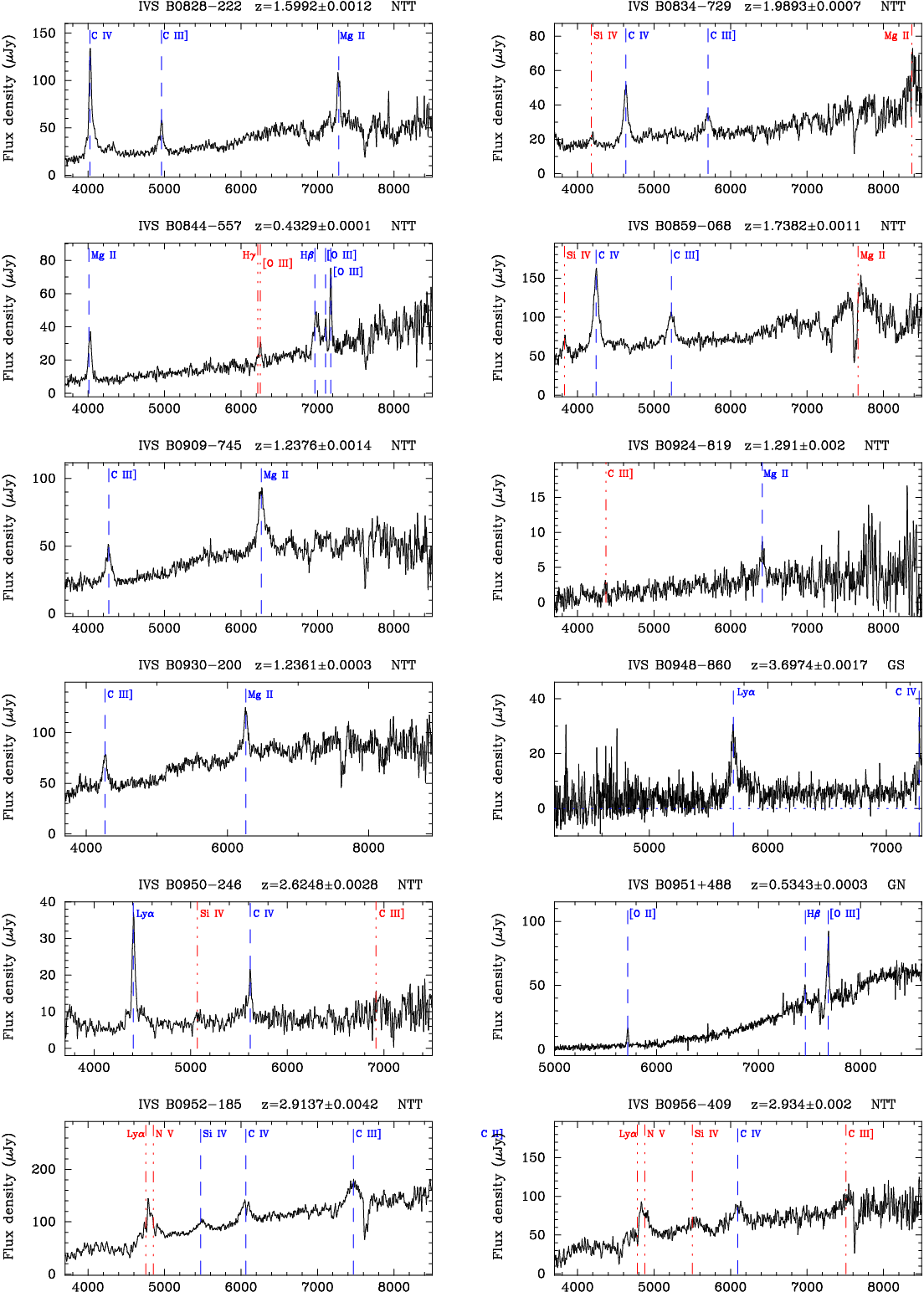}
\caption{}
\end{figure*}

\begin{figure*}
\centering
\figurenum{1 (continued)}
\includegraphics[width=0.85\textwidth]{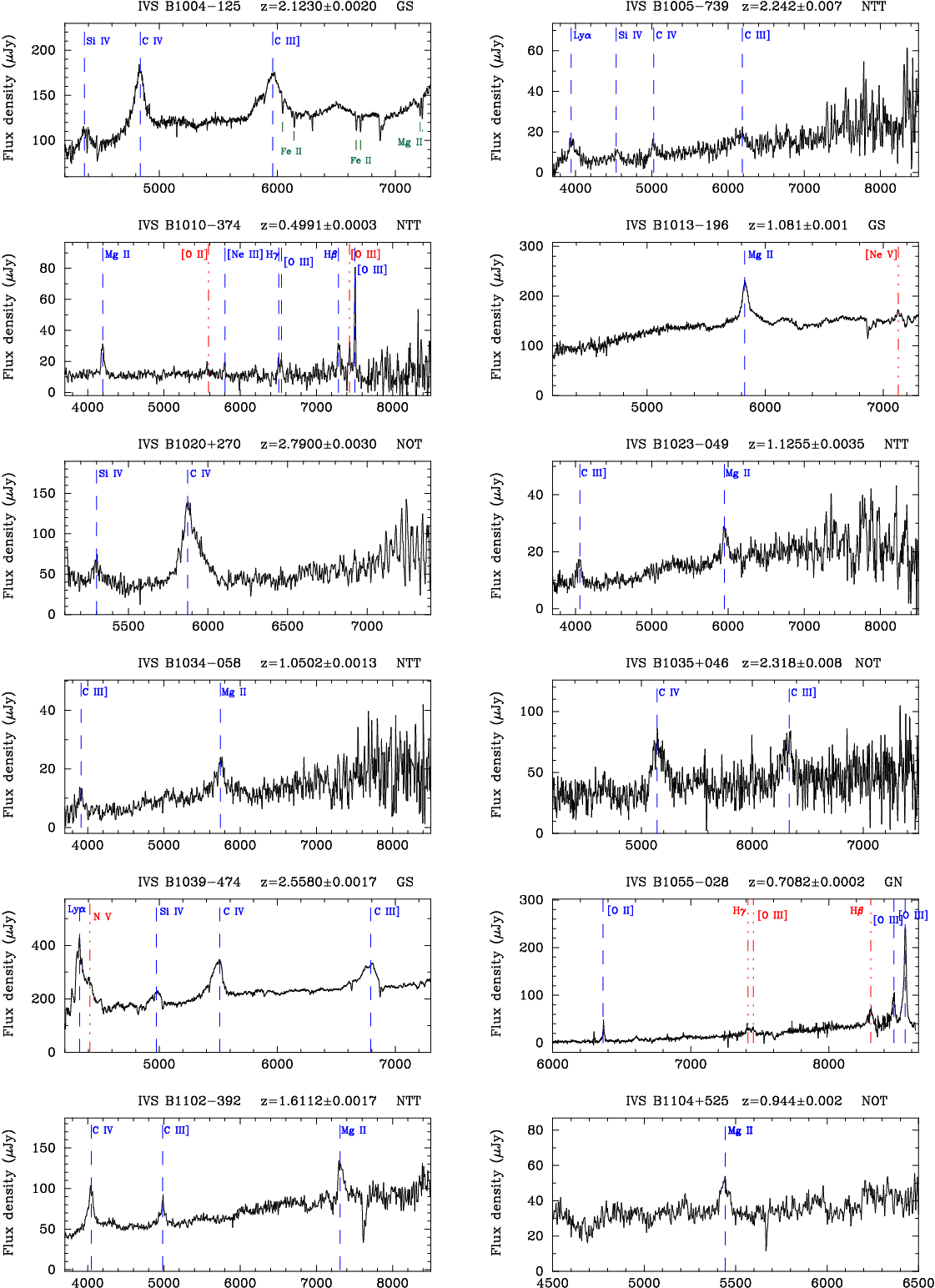}
\caption{}
\end{figure*}

\begin{figure*}
\centering
\figurenum{1 (continued)}
\includegraphics[width=0.85\textwidth]{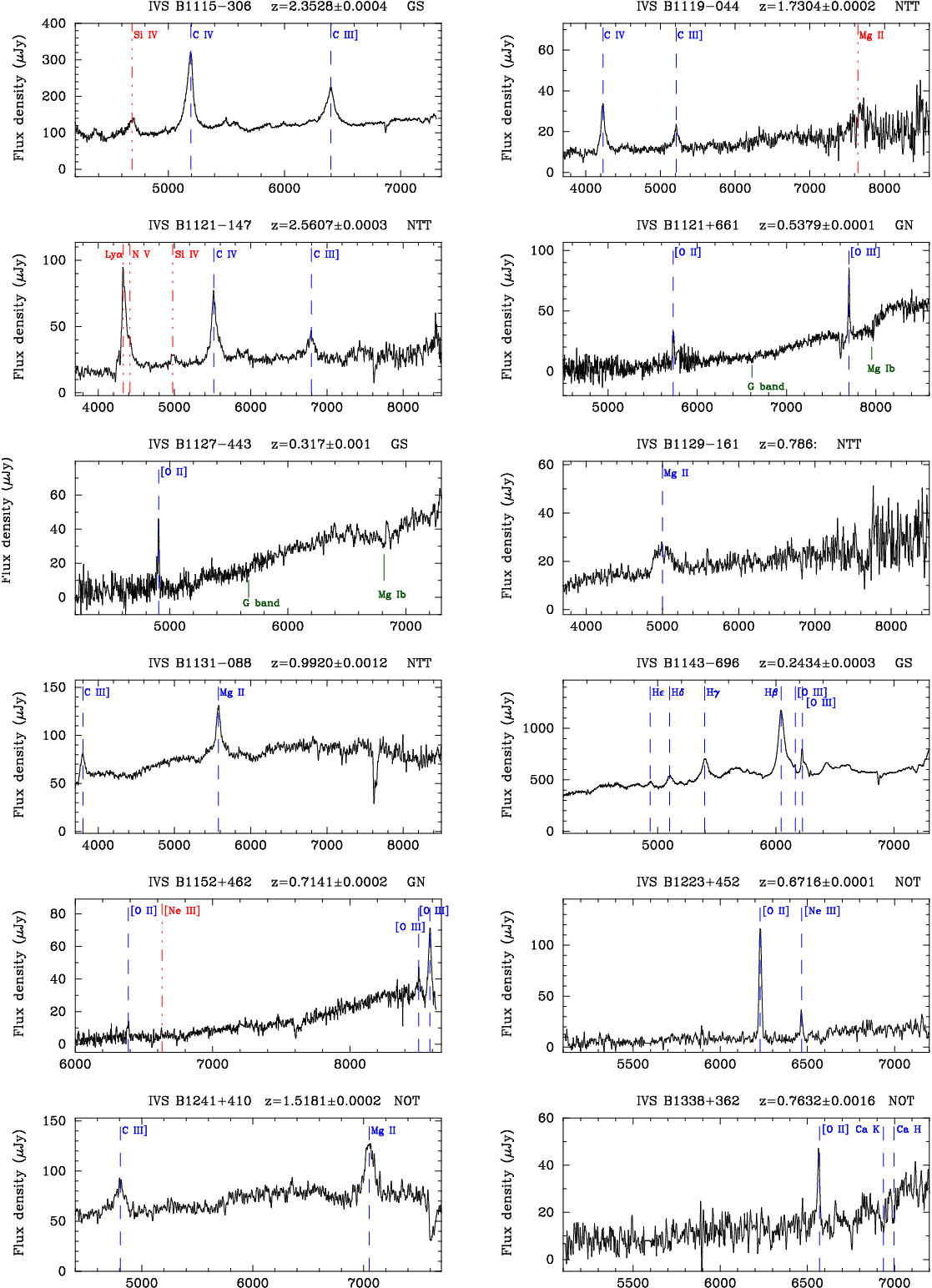}
\caption{}
\end{figure*}

\begin{figure*}
\centering
\figurenum{1 (continued)}
\includegraphics[width=0.85\textwidth]{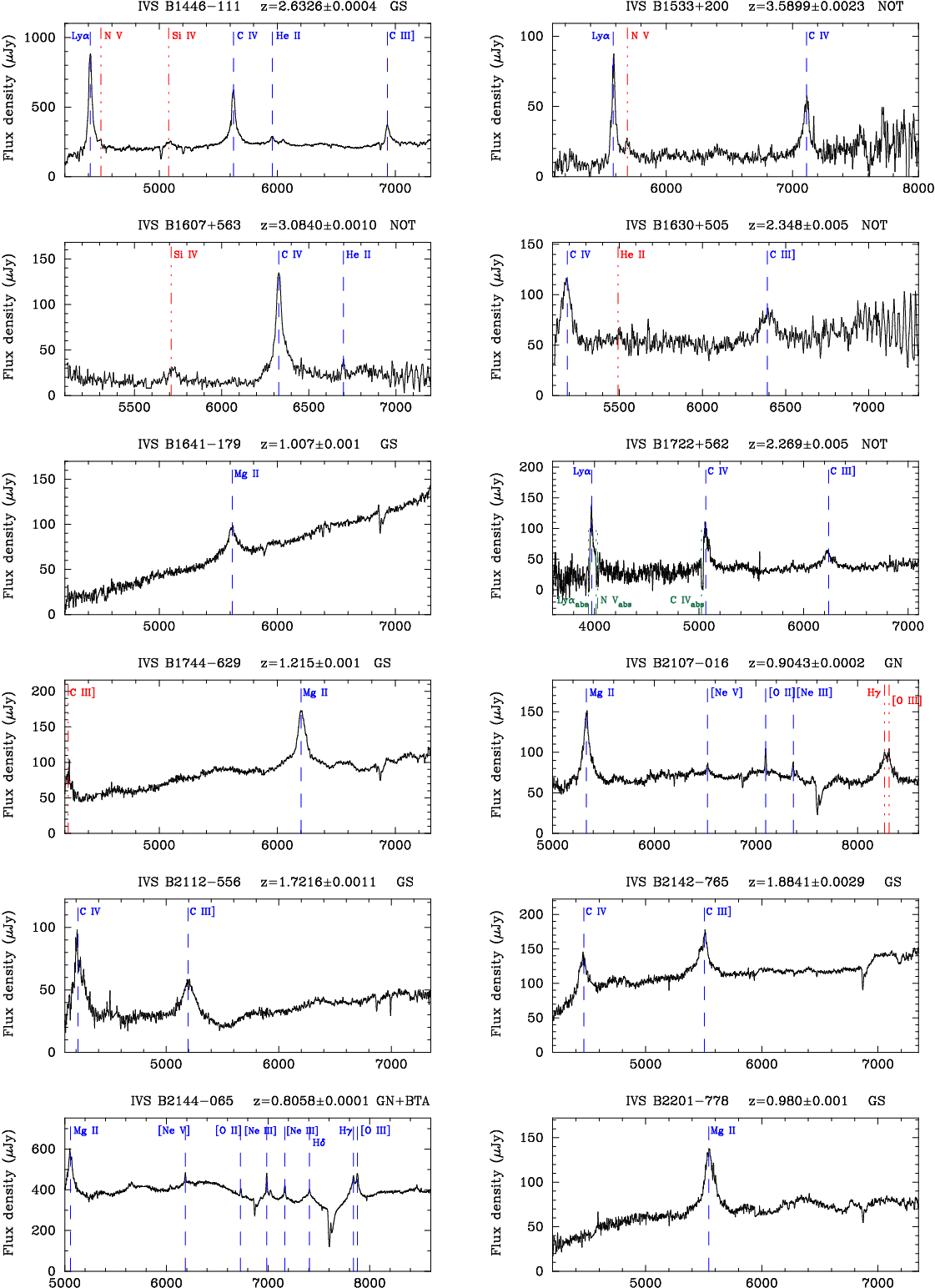}
\caption{}
\end{figure*}

\begin{figure*}
\centering
\figurenum{1 (continued)}
\includegraphics[width=0.85\textwidth]{fig1j.eps}
\caption{}
\end{figure*}

\begin{figure*}
\centering
\figurenum{2}
\includegraphics[width=0.85\textwidth]{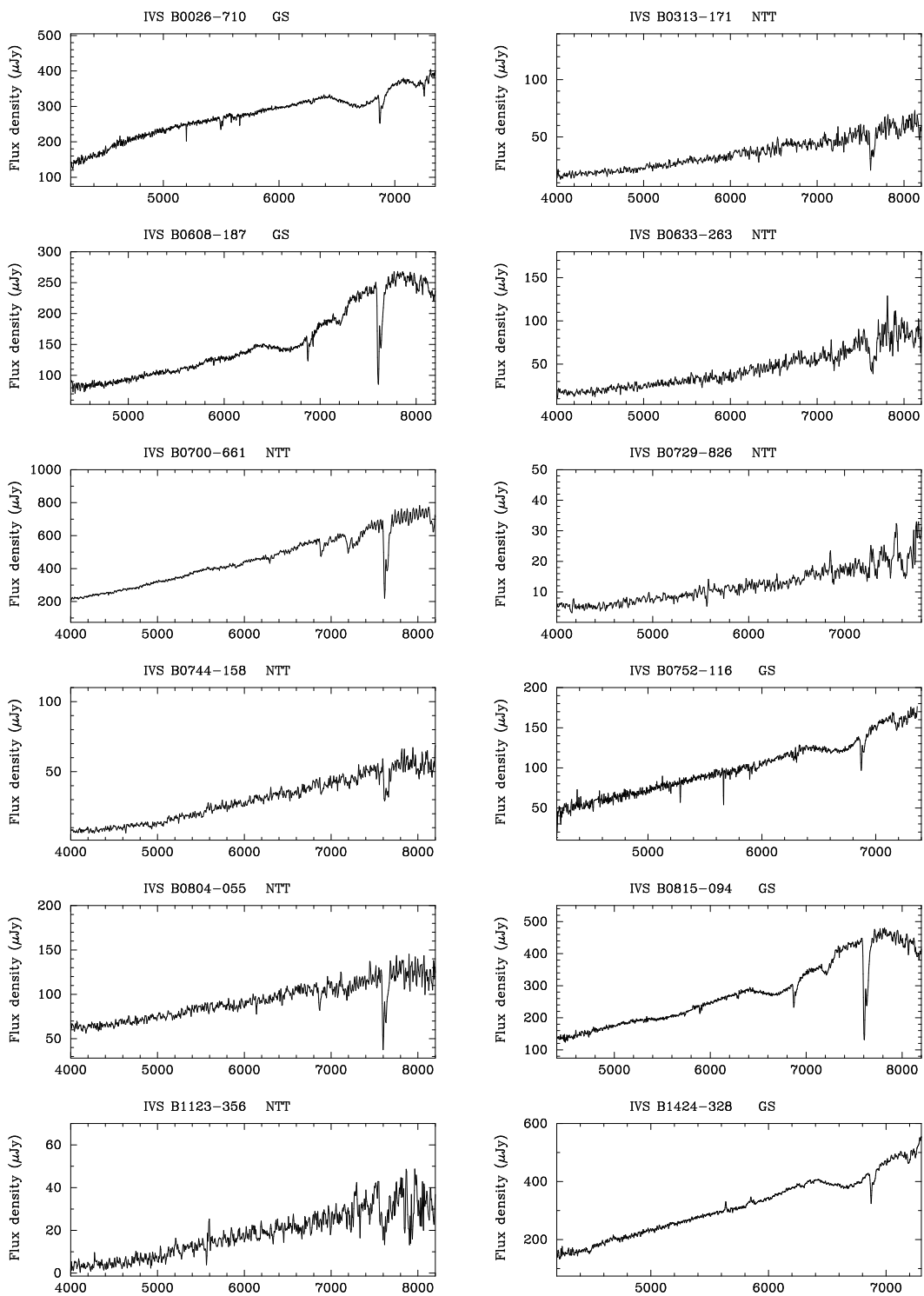}
\caption{Spectra of 19 probable BL Lac objects from our observations, 
classified on the basis of their featureless spectra; see Table\ 
\ref{BLL} for ICRF2 positions. \label{BLLspectra}} \end{figure*} 

\begin{figure*}
\centering
\figurenum{2 (continued)}
\includegraphics[width=0.85\textwidth]{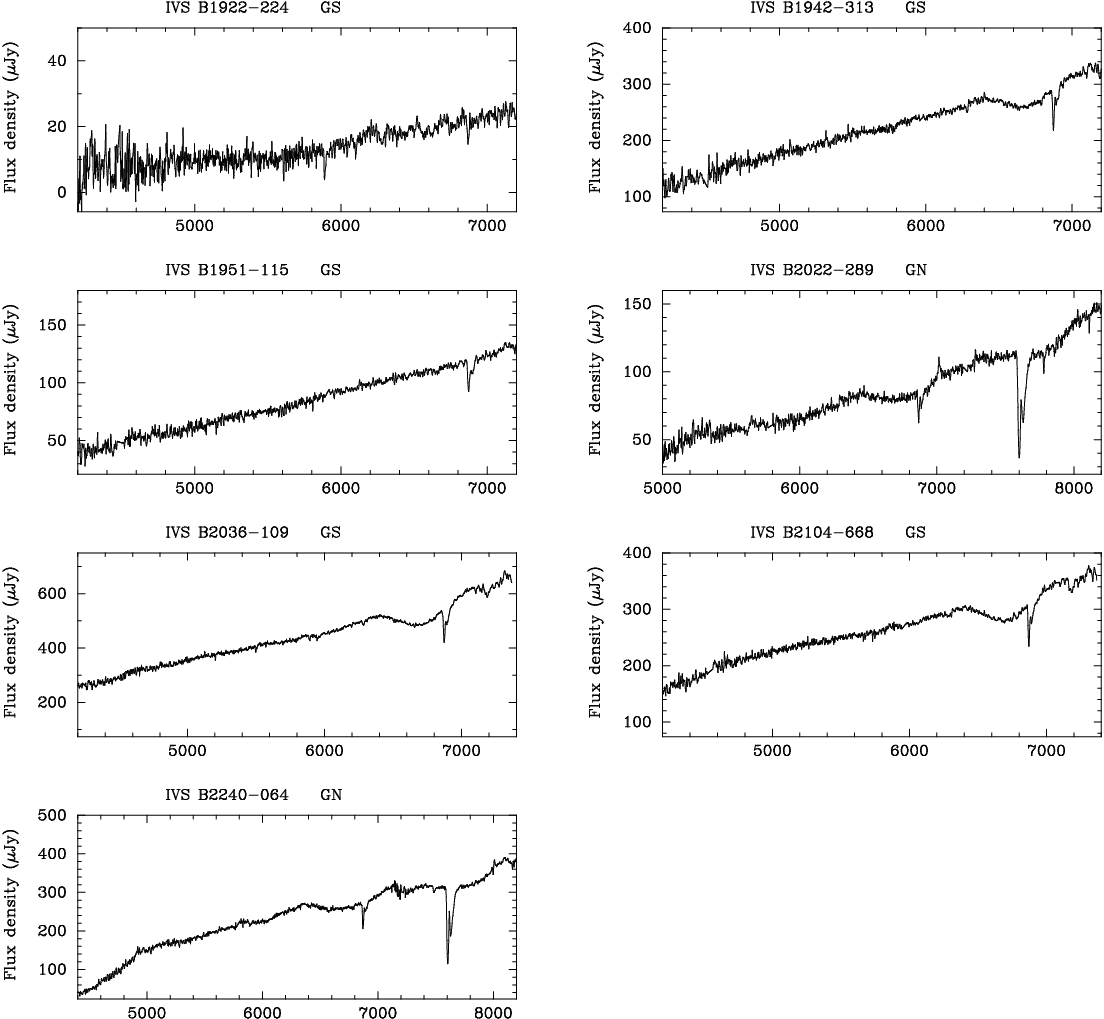}
\caption{}
\end{figure*}

\end{document}